\def\eeq{\end{equation}}

\def\beq{\begin{equation}}

\def\bea{\begin{eqnarray}}

\def\eea{\end{eqnarray}}

\documentstyle[aps]{revtex}

\begin{document}

\title{Generalized quantal distribution functions within factorization
approach: Some general results for bosons and fermions}

\author{Diego F. Torres$^{1,2,}$\thanks{dtorres@venus.fisica.unlp.edu.ar}
and
U\v{g}ur T{\i}rnakl{\i}$^{3,}$\thanks{
tirnakli@sci.ege.edu.tr}}

\address{$^1$ Departamento de F\'{\i}sica, Universidad Nacional de La Plata,
C.C. 67, 1900, La Plata,  Argentina\\
$^2$Astronomy Centre, CPES, University of Sussex, Falmer,
Brighton BN1 9QJ, United Kingdom\\
$^3$ Department of Physics, Faculty of Science, Ege University
35100 Izmir-Turkey}

\maketitle

%%%%%%%%%%%%%%%%%%%%%%%%%%%%%%%%%%%%%%%%
\begin{abstract}
The generalized quantal distribution functions are investigated concerning
systems of non-interacting bosons and fermions. The formulae for the
number of particles and energy are presented and applications to the
Chandrasekhar limit of white dwarfs stars and to the Bose-Einstein
condensation are commented.

\noindent
{\it PACS Number(s):} 05.20.-y, 05.30.Jp, 05.30.Fk

\end{abstract}
%%%%%%%%%%%%%%%%%%%%%%%%%%%%%%%%%%%%%%%%

\newpage

%\vspace{1.5cm}

\section{Introduction}
Since 1988, a non-extensive formalism of statistical mechanics
\cite{tsallis1} has been developed and up to
recent years, many works have been
devoted to show the robustness and usefulness
of this approach. We believe it
is robust in the sense it allows generalizations of a
variety of fundamental concepts of statistical thermodynamics
\cite{concepts}, such that it avoid to enter in severe contradiction
with well established facts. In addition, we believe it is
useful in the sense it provides a theoretical basis and
relevant explanation of some experimental and observational situations
\cite{verify}, where Boltzmann-Gibbs statistics fails. That is to say, where
Boltzmann-Gibbs statistical functions diverge and  do not yield any physical
prediction.

However, this formalism, unlike Boltzmann-Gibbs', has a non-zero set of free
parameters, here represented by $q$ ($q\in\Re$). This
unique parameter controls the degree of the
nonextensivity of the system in consideration. (At this point, it is worth
noting that the formalism includes the standard, extensive,
statistics as a special case for the value of $q=1$ and all
expressions derived within this non-extensive framework give the results of
Boltzmann-Gibbs statistics in the $q\rightarrow 1$ limit.)
Only after 1995, some
works started to address the long standing puzzle of understanding the
physical meaning of $q$.
Amongst the works related to this topic, two main streams started to
become more apparent. On one side,
there are attempts on the study of conservative
\cite{hamiltonian} and
dynamical systems, more precisely, dissipative systems with both, low
\cite{low} and high \cite{high} dimensions.
On the other, there have been some efforts of estimating
bounds upon $q$ in measurable physical systems. The
physical applications studied so far can be enumerated as follows: the
microwave background radiation \cite{t-radiation,u-radiation}, the
Stefan-Boltzmann constant \cite{p-stefan,u-stefan}, the early Universe
\cite{d-early,u-early} and the primordial neutron to baryon ratio in a
cosmological expanding background \cite{diego}.
As a final note, a recent letter by Alemany \cite{ALEMANY}
explored the definition of a new
{\it fractal canonical ensemble}, associated with the parameter $q$. However,
first estimations for the universe as a whole yielded to values of $q$ bigger
than those allowed by nucleosynthesis \cite{diego}.

In all these works, two
different approaches for the quantal distribution functions have been
used: a closed analytical form for them is still lacking.
The asymptotic approach of the kind
$\beta (1-q)\rightarrow 0$ of Tsallis et al. \cite{t-radiation} was
used in Refs. \cite{t-radiation,p-stefan,d-early,diego} and
also in  some other applications such as the study
of Bose-Einstein condensation in a fractal space
\cite{curilef}, the specific heat of $^4\!$He
\cite{curilefpapa} and the thermalization of an electron-phonon system
\cite{koponen}. The second approach for the generalized
distribution functions has been proposed by B\"{u}y\"{u}kk{\i}l{\i}\c{c} et
al. \cite{buyukkilic}, which we refer to as {\it Factorization Approach}. This last
term is justified in that
it is generically based on the factorization of the
generalized grand canonical distribution and the concomitant generalized
partition function as if they were extensive quantities. However, although
the generalized distribution functions within factorization approach have
been derived before the asymptotic approach, they have not been prefered to
be used in physical applications, mainly due to a work by Pennini et al.
\cite{pennini}. There, the authors claimed that the quality of the results of
the factorization approach deteriorates as the number of the particles of
the system increases.
But very recently, contrary to the previous belief, new
results by Wang and L\'{e} M\'{e}haut\'{e} \cite{wang},  favoured the
factorization approach and showed clearly that there exists a temperature
interval where the ignorance of the approximation is significant, but
otherwise, the results of this approach can be used with
confidence no matter the number of particles.
In fact, they showed that, for a macroscopic system (i.e., with $\sim
10^{23}$ particles) having two states with a small energy interval of about
$10^{10}$ Hz, this forbidden temperature zone is very narrow and
situated at extremely high temperatures ($\sim 10^{20}$ K) and as the number
of levels increases the forbidden temperature zone shifts to higher
temperatures. Therefore the results of the
factorization approach can be used at
low temperatures up to $10^{20}$ K without any ignorance for any physical
system under consideration \cite{wang,wang2,wang3}.
As it was
expected, the study of Wang and L\'{e} M\'{e}haut\'{e} started to accelerate
new attempts of applying these distribution functions to physical systems in
order to
estimate some alternative bounds upon $q$ \cite{u-stefan,u-early}; being this
approach more handable than the one advanced by Tsallis et al.
Although, after Wang and L\'{e} M\'{e}haut\'{e} \cite{wang}, some physical
systems have been worked out with the help of the generalized distribution
functions of the factorization approach, general expressions for
magnitudes of fermions and bosons are still lacking.

Indeed, the possible
need for a nonextensive formalism of thermostatistics is clear
for a long time in gravitation \cite{gravity}, magnetic systems
\cite{magnet}, L\'{e}vy-like diffusions \cite{levy-like}, some
surface-tension problems \cite{surface}, etc., since Boltzmann-Gibbs
formalism is known to fail whenever the physical system under consideration
includes (i) long-range interactions and/or (ii) long-memory effects and/or
(iii) the system evolves in a multifractal-like space-time. The last
property can be understood as the system evolves in a porous-like medium
where the properties of the multifractal structures govern it.

In the light of the facts stated above, we address the study
of the generalized quantal distribution functions of fermions and bosons
in a non-Euclidean, fractal-like space-time and to obtain general results for
magnitudes, such as the number of particles and the internal energy
that could be used in general for any application to generalized systems.
Once these general formulae be established, we proceed further investigating
how some simple constructs of the statistical theory are affected by
the change of distribution functions,
for instance, the Chandrasekhar mass and the Bose-Einstein
condensation. A similar work studying the Bose-Einstein condensation
was recently done, as commented above,  by Curilef \cite{curilef} and our
results are compatible  in this case.
Finally, it is worth noting that, for non-interacting fermions and bosons,
the information about the fractal structure is kept in the nonextensivity
parameter $q$, which is by now verified to be connected with the fractal
dimension and the multifractal singularity spectrum \cite{low}. In a sense, we
shall study if any correction to the usual number of particles
and energy arises due to
the change of the distribution functions.

\section{The fermion case}

Let us start by writing down the generalized distribution function for
fermions, up to $(1-q)$ order. It is given by \cite{buyukkilic},

\beq
n_{q[fermions]} = \frac{1}{e^{\beta(\epsilon-\mu)} + 1} +
\frac{q-1}{2} \frac{\left[\beta(\epsilon-\mu)\right]^2
e^{\beta(\epsilon-\mu)}}{\left(e^{\beta(\epsilon-\mu)} + 1\right)^2} ,
\eeq
where $\beta=1/kT$, $\epsilon$ is the energy level and $\mu$ is
the chemical potential.

We are interested in solving the following integrals. For the number of
particles,

\beq
\label{total}
<N>=\frac {4\pi V (2m)^{3/2}}{h^3} \left[ \int_0^\infty d\epsilon
\frac{\epsilon^{1/2}}{e^{\beta (\epsilon - \mu)} +1}
+ \frac {q-1}{2} \int_0^\infty d\epsilon \frac{ (\beta(\epsilon - \mu ))^2
e^{\beta (\epsilon - \mu)} \epsilon^{1/2}}{(e^{\beta (\epsilon - \mu)} +1)^2}
\right]
\eeq
and a similar one, replacing the factors $\epsilon^{1/2}$ for $\epsilon^{3/2}$,
for the energy $<U>$.
Note that we have assumed the same pre-factor for both terms.
This can be justified if one thinks of
this factor as coming from microscopic quantum
considerations and a change of variables, from momentum
to energy, inside the integrals. See for instance, section 2.4 of Ref.
\cite{PATHRIA}.

Then, we can split $<N>$ into two parts. Firstly we have the usual expression,
given by the first term in the rhs of (\ref{total}),

\bea
Usual=\frac {4\pi V (2m)^{3/2}}{h^3}  \int_0^\infty d\epsilon
\frac{\epsilon^{1/2}}{e^{\beta (\epsilon - \mu)} +1}
=\frac {8\pi V (2m\mu)^{3/2}}{3h^3} \nonumber \\
\hspace{3cm}\left[ 1 + \frac 18 \left( \frac{\pi^2}{\beta \mu}
\right)^2 + \frac{7}{640}
\left(\frac{\pi}{\beta \mu} \right)^4 + \ldots \right].
\eea
And we also have a second term,

\beq
I_q=\frac {4\pi V (2m)^{3/2}}{h^3}  \frac {q-1}{2}
\int_0^\infty d\epsilon \frac{ (\beta(\epsilon - \mu ))^2
e^{\beta (\epsilon - \mu)} \epsilon^{1/2}}{(e^{\beta (\epsilon -
\mu)} +1)^2} =
\frac {4\pi V (2m)^{3/2}}{h^3}\frac {q-1}{2} I  .
\eeq
$I$ can be recasted using the following dimensionless variables:

\beq
x=\beta \epsilon, \hspace{2cm} \psi=\beta \mu.
\eeq
This enables us to write

\beq
I= \beta^{-3/2} \int_0^\infty \frac{(x-\psi)^2 e^{x-\psi} x^{1/2}}
{(e^{x-\psi} +1)^2} dx =
\beta^{-3/2} I_x.
\eeq
In order to solve this last $I_x$ we apply the following trick.
We know that integrals of the form

\beq
\label{int}
\int_0^\infty \frac{f(x)}{e^{x-\psi} +1 } dx
\eeq
have a solution,

\beq
\label{serie}
\int_0^\psi f(x) dx + \frac{\pi^2}{6} \frac {df}{dx} +
\frac {7 \pi^4}{360} \frac{d^3 f}{dx^3} + \frac {31 \pi^6}{15120}
\frac{d^5f}{dx^5} + \ldots ,
\eeq
where the derivatives have to be evaluated in $x=\psi$
\cite{PATHRIA}. In fact, this was already used to write down
the solution for the usual term of (\ref{total}).
The idea is then to introduce an extra parameter, $l$, in the
integral (\ref{int})
and consider,

\beq
\label{int-l}
\int_0^\infty \frac{f(x)}{e^{x-l\psi} +1 } dx .
\eeq
Here, it is important to stress that $f(x)$ does not depend on $l$. If we
derive inside this integral with respect
to $l$, the result will be

 \beq
\label{int-f}
\int_0^\infty \frac{f(x) \psi e^{x-l\psi} }{(e^{x-l\psi} +1)^2 } dx ,
\eeq
and if $l=1$, it can reproduce our integral $I_x$ iff the function $f(x)$ is
given by,

\beq
\label{def}
f(x)= \frac{(x-\psi)^2 }{\psi} x^{1/2} .
\eeq
Thus, using (\ref{def}), the solution to $I_x$ is,

\beq
I_x= \left[ \frac {d}{dl} \int_0^\infty \frac{f(x)}{(e^{x-l\psi} +1) } dx
\right]_{l=1}.
\eeq
We have to be very careful in using the correct formula to get
the integral in $f(x)$:  we should
redefine $\tilde \psi=l\psi$ and apply then (\ref{serie}).
So, the mechanism is, considering $f(x)$ as in (\ref{def}),
compute the integral
(\ref{int-l}),
derive the result with respect to $l$ and finally evaluate in $l=1$.
That yields, as we have seen, the integral we need.\footnote{A more transparent
way of solving these integrals (see acknowledgments) is to take the
integral (6) as
\beq
\frac{\partial}{\partial \psi} \int_0^{\infty} \frac{x^{5/2}}
{e^{x-\psi} +1} dx - 2 \psi \frac{\partial}{\partial \psi} \int_0^{\infty}
\frac{x^{3/2}}{e^{x-\psi} +1} dx + \psi^2 \frac{\partial}{\partial\psi}
\int_0^{\infty} \frac{x^{1/2}} {e^{x-\psi} +1} dx
\eeq
and apply expansion (8) to each part. As we see, in fact there is no need
to introduce such an
extra parameter $l$, and it only stands as a mathematical
trick in a particular form of solving the integrals. Numerical results
for the coefficients obtained with both methods are the same. We warn the
reader not to be distracted for such a trick.}

Let treat each of the term in (\ref{serie}) separately. The first term is
obtained inmediately:

\beq
\int_0^{\tilde \psi} \frac{(x-\psi)^2 }{\psi} x^{1/2} dx = \frac 1\psi
 \int_0^{\tilde \psi} (x^2 -2 x \psi + \psi^2) x^{1/2} dx,
\eeq
which finally leads to,

\beq
1^{st}\;\;term\;=\;\frac 1\psi \left[ \frac{\tilde \psi^{7/2}}{7/2} - 2 \psi
 \frac{\tilde \psi^{5/2}}{5/2} + \psi^2  \frac{\tilde \psi^{3/2}}{3/2}
\right].
\eeq
Now we have to derive with respect to $l$
(recall that $\tilde \psi =  l \psi$), obtaining

\beq
\frac{d\; 1^{st}\;\;term}{dl} = \frac 1\psi \left[ (7/2) l^{5/2}
\frac{ \psi^{7/2}}{(7/2)} - 2
(5/2) l^{3/2} \frac{ \psi^{7/2}}{(5/2)} + (3/2) l^{1/2}
\frac{\psi^{7/2}}{(3/2)} \right] ,
\eeq
and evaluate at $l=1$. The result is,

\beq
\left[ \frac{d\; 1^{st}\;\;term}{dl} \right]_{l=1}=0.
\eeq
The second term in (\ref{serie}) is obtained as,

\beq
2^{nd}\;\;term\;= \;
\frac{\pi^2}{6} \left[ 2\frac{(x-\psi)}{\psi} x^{1/2} +
\frac{(x-\psi)^2}{\psi}
\frac 12
x^{-1/2} \right]_{\tilde \psi},
\eeq
which finally yields to,

\beq
2^{nd}\;\;term\;= \;
\frac{\pi^2}{6} \left[ 2\frac{(l\psi-\psi)}{\psi} (l\psi)^{1/2} +
\frac{(l\psi-\psi)^2}{\psi}
\frac 12
(l\psi)^{-1/2} \right].
\eeq
The derivative with respect to $l$ is,

\bea
\frac{6}{\pi^2} \frac{d\; 2^{nd}\;\;term}{dl} =
 2 (l\psi)^{1/2} + 2(l\psi-\psi) \frac 12  (l\psi)^{-1/2}
+ 2 (l \psi -\psi)
\frac 12 (l\psi)^{-1/2} \nonumber \\
\;\;\;\;\;\;+ (l\psi-\psi)^2 \frac 12 l^{-3/2} \frac{-1}{2}  \psi^{-3/2} .
\eea
Evaluating in $l=1$ we obtain,

\beq
\left[ \frac{d\; 2^{nd}\;\;term}{dl} \right]_{l=1}=
\frac{\pi^2}{3} \psi^{1/2}.
\eeq
One can do the same with the third term. In this case we need to
compute the third derivate. It is given by,

\beq
f^{iii}=
\frac 3\psi x^{-1/2}
- \frac{3}{2} (x-\psi) \frac{x^{-3/2}}{\psi} + (x- \psi)^2  \frac 38
\frac {x^{-5/2}}{\psi}.
\eeq
Computing this last for $x=\tilde \psi$ and deriving with respect to $l$
we get,

\beq
\frac{df^{iii}}{dl} =
-3  (l\psi)^{-3/2}   +
3 (l\psi- \psi) (l\psi )^{-5/2} +
(l\psi- \psi)^2 \left(\frac{-15}{16}\right)  (l\psi)^{-7/2}.
\eeq
And finally making $l=1$, the correction is found to be,

\beq
\left[ \frac{df^{iii}}{dl} \right]_{l=1} =
-3  (\psi)^{-3/2}.
\eeq

Now we can go back and collect the results for $I$, which results,

\beq
I=\beta^{-3/2} I_x=\beta^{-3/2} \left( 0+ \frac{\pi^2}{3} \psi^{1/2} -
\frac{7 \pi^4}{120} \psi^{-3/2} + \ldots
\right),
\eeq
and the final expression for $<N>$ is then,

\bea
<N> \frac{3h^3}{8\pi V (2m\mu)^{3/2}}=
 \left( 1 + \frac 18 \left( \frac{\pi^2}{\beta \mu} \right)^2 + \frac{7}{640}
\left(\frac{\pi}{\beta \mu} \right)^4 + \ldots \right) \nonumber \\
 +\frac {q-1}{2} \left( 0 + \frac{\pi^2}{2}
\frac{1}{\beta \mu} - \frac{21 \pi^4}{240} \frac{1}{(\beta \mu)^3} +
\ldots \right).
\eea
This final expression represents the correction terms due to non-extensivity
that arise for $<N>$.

We can do exactly the same for the energy $<U>$. $I_q$ is given now by,

\beq
I_q=\frac {4\pi V (2m)^{3/2}}{h^3}\frac {q-1}{2} \beta^{-5/2} I_x ,
\eeq
where $I_x$ is,

\beq
I_x= \int_0^\infty \frac{ (x-\psi)^2 e^{x-\psi} x^{3/2}}{(e^{x-\psi} +1 )^2}
dx .
\eeq
We apply again the same trick,
in this case, with a function $g(x)= (1/\psi) (x-\psi)^2 x^{3/2}$.
As before,

\beq
I_x= \left[ \frac {d}{dl} \int_0^\infty \frac{g(x)}{(e^{x-l\psi} +1) } dx
\right]_{l=1}.
\eeq
The first term in the
serie (\ref{serie}), is again equal to zero, as can be directly verified by
computing
the integral for $g$, deriving the result with respect to $l$ and evaluating
at $l=1$. The second
term needs

\beq
\frac{dg}{dx}= \frac{1}{\psi} \left[ 2(x-\psi) x^{3/2} + (x-\psi)^2
\frac 32 x^{1/2} \right].
\eeq
Derivating this with respect to $l$ and evaluating at $l=1$ we obtain for the
correction,

\beq
\frac{\pi^2}{3} \psi^{3/2}.
\eeq
The third term needs the third derivative of $g$. This is found to be,

\beq
\frac 1\psi \left[ 9 x^{1/2} + x^{-1/2} (x-\psi) \frac 92 - (x-\psi)^2
\frac 38 x^{-3/2} \right].
\eeq
Again deriving with respect to $l$ and evaluating at $l=1$, it yields the
following correction

\beq
\frac {7\pi^4}{360} 9 \psi^{-1/2}.
\eeq
Collecting the results, we get for $<U>$,

\bea
<U>  \frac{5h^3}{8\pi V (2m)^{3/2} \mu^{5/2}} =
 \left( 1 + \frac 58 \left( \frac{\pi^2}{\beta \mu} \right)^2 -
 \frac{7}{384}
\left(\frac{\pi}{\beta \mu} \right)^4 + \ldots \right) \nonumber \\
\;\;\;\;\;\;+ \frac {q-1}{2} \left( 0 + \frac{5\pi^2}{6}
\frac{1}{\beta \mu} + \frac{315 \pi^4}{720} \frac{1}{(\beta \mu)^3} +
\ldots \right).
\eea
As above, this expression represents the correction due to non-extensivity
that arise for the energy.

It is important to note that the {\it $T \rightarrow 0$ limit is attained
without any correction} thus
suggesting that {\it non-extensivity can hardly have any role for very
low temperatures.}
One can now study  particular physical systems. For instance, one can
inmediately realize that the Chandrasekhar mass will
be not modified at all in passing to a non-extensive
context, at least up to order  $(1-q)$.
This is simple because the Chandresekhar mass is a construct that arises at
$T=0$ where both corrections, to $<N>$ and to $<U>$, are exactly zero.
However, if non-extensive statistics is concerned in the study of stellar
structures, corrections will unavoidably arise.

\section{The boson case}

Now we analize the boson generalized distribution function, namely,

\beq
\label{bos-n}
n_{q[bosons]} = \frac{1}{e^{\beta(\epsilon-\mu)} - 1} +
\frac{q-1}{2} \frac{\left[\beta(\epsilon-\mu)\right]^2
e^{\beta(\epsilon-\mu)}}{\left(e^{\beta(\epsilon-\mu)} - 1\right)^2} ,
\eeq
where again $\beta=1/kT$, $\epsilon$ is the energy level, $\mu$ is the
chemical potential. Then, using equation  (\ref{bos-n}), one can obtain
the average number of particles as

\beq
\label{bos-int}
\frac{\left(N\right)_q}{V} = \frac{2\pi (2m)^{3/2}}{h^3}
\left[\int_0^{\infty}\frac{\epsilon^{1/2} d\epsilon}
{e^{\beta(\epsilon-\mu)} -1} + \frac{q-1}{2}
\int_0^{\infty} \frac{\beta^2(\epsilon-\mu)^2 e^{\beta(\epsilon-\mu)}
\epsilon^{1/2} d\epsilon}{\left(e^{\beta(\epsilon-\mu)} - 1\right)^2}
\right].
\eeq
As in the standard case, we have to separate the state with $\epsilon =0$,
which has zero weight in the integral (\ref{bos-int}). For this level of
energy, the distribution function yields,

\beq
N_q(\epsilon=0)=\frac{z}{1-z}\left[1+\frac{q-1}{2}\frac{(\ln z)^2}{1-z}\right]
\eeq
where $z=e^{\beta\mu}$ is the fugacity of the gas. In the case $z \ll 1$, the
correction term goes to zero as $z(\ln  z)^2$ does. If $z\rightarrow 1$, the
correction goes to 1 as $z$, and results are neglectable in comparison with the
first, diverging, term. This can be seen in Fig. 1.

Substitution
of $x=\beta\epsilon$ and $\psi=\beta\mu$ in equation  (\ref{bos-int}) yields,

\beq
\label{bos-int-2}
\frac{\left(N_e\right)_q}{V} = \frac{2\pi (2mk)^{3/2}}{h^3} T^{3/2}
\left[ I_1 +\frac{q-1}{2} I_q \right]
\eeq
where $\left(N_e\right)_q$ stands for the number of particles in
the excited states ($\epsilon\neq0$), and $I_1$ and $I_q$ are defined to be

\beq
I_1 = \int_0^{\infty}\frac{x^{1/2} dx}{e^{x-\psi} -1}, \;\;\;\;\;  \;\;\;\;\;
I_q = \int_0^{\infty}\frac{(x-\psi)^2 e^{x-\psi} x^{1/2} dx}
{\left(e^{x-\psi} -1\right)^2} .
\eeq
The integral $I_1$ is the one which appears in the standard,
$q=1$ case of the boson gas and therefore the solution of it is the known one
\cite{PATHRIA},

\beq
I_1 = \Gamma(3/2) g_{3/2}(z) ,
\eeq
where $g_{n}(z)=\frac{1}{\Gamma (n)} \int_0^\infty \frac{x^{n-1}}
{z^{-1}e^x -1} dx
\simeq
\sum_{s=1}^{\infty}\left(z^s/s^n\right)$ for small $z$ and $\Gamma$ is the
usual Gamma function, $\Gamma(n)=\int_0^\infty e^{-x} x^{n-1} dx$.
For $0\leq z \leq 1$ and $\forall n, n>1$, the functions $g_n(z)$ are bounded
by the Riemann zeta functions, which yields for all $z$ values of interest,

\beq
I_1 \leq \Gamma(3/2) \zeta(3/2) .
\eeq
On the other hand, the integral $I_q$ is again
somewhat cumbersome to
solve, but after some algebra, similar to what we did in the previous section,
we managed to find the analytical solution as

\beq
I_q = \Gamma(7/2) g_{5/2}(z) - 2\psi \Gamma(5/2) g_{3/2}(z) +
\Gamma(3/2) \psi^2 g_{1/2}(z) .
\eeq
To get the previous result one has to take into account the known
relationship between the $g_n(z)$ and its derivatives,

\beq
g_{n-1}(z)=z\frac{\partial}{\partial z} g_n(z)= \frac{\partial}{\partial
(\ln z)}g_n(z).
\eeq
Recalling again the definitions $\psi$ and
$z$, we can write $\psi=\ln z$.
Putting these solutions of the integrals in equation  (\ref{bos-int-2}),
one can obtain

\bea
\frac{\left(N_e\right)_q}{V} = \frac{2\pi (2mk)^{3/2}}{h^3} T^{3/2}\times
\;\;\;\;\;\;\;\;\;\;\;\;\;\;\;\;\;\;\;\;\;\;
 \nonumber
\\
\left\{\Gamma(3/2) g_{3/2}(z) + \frac{q-1}{2} \left[
\Gamma(7/2) g_{5/2}(z) - 2 \Gamma(5/2) \ln z  g_{3/2}(z) +
\Gamma(3/2) (\ln z)^2 g_{1/2}(z) \right] \right\},
\eea
which is the $q$-dependent solution for the number of particles
in the excited states of boson systems, including the standard case as a
special one if   $q=1$.

Like in the standard case,
 let us study each term of the correction in a  separate fashion
 in order to find the bounded
form of them.
The first term is similar to
$I_1$ and hence it is bounded by $\Gamma(7/2)\zeta(5/2)$. The second and the
third terms are different from the first one in the sense that they do not
take their largest values at $z=1$, but instead, both of them tend to $0$
when $z\rightarrow 0$ and $z\rightarrow 1$.
This behavior can be seen in Fig. 2.
The largest values  are situated at $z_{max1}\simeq 0.447$ for the second term
and $z_{max2}\simeq 0.175$ for the third.

Consequently, if all these bounds are put together, then the total number of
particles in all excited states is also bounded by,

\bea
\left(N_e\right)_q \leq V \frac{2\pi (2mk)^{3/2}}{h^3} T^{3/2}
\left\{
\Gamma(3/2)\zeta(3/2) + \frac{q-1}{2} \times
\;\;\;\;\;\;\;\;\;\;\;\;\;\;\;\;\;\;\;\;\;\;\;\;\;
 \right. \nonumber\\
 \left.
\left[\Gamma(7/2)\zeta(5/2) - 2\Gamma(5/2) \ln (0.447) g_{3/2}(0.447)
+ \Gamma(3/2) (\ln (0.175))^2 g_{1/2}(0.175)
 \right] \right\}
\end{eqnarray}
which gives

\beq
\left(N_e\right)_q \leq V \frac{2\pi (2mk)^{3/2}}{h^3} T^{3/2}
\left\{2.315 + (q-1) 3.079 \right\} .
\eeq

If we now concentrate on Bose-Einstein condensation, then the condition for
the appearance of it can be expressed as

\beq
N > \left(N_e\right)_q .
\eeq
Alternatively, with constant $N$ and $V$, this condition can be recasted in the
form,

\beq
T < \left(T_c\right)_q =\frac{h^2}{\left(2\pi\right)^{2/3} 2mk}
\left\{\frac{N}{V\left[2.315+(q-1)3.079\right]}\right\}^{2/3},
\eeq
or up to order 1 in $(q-1)$,

\beq
T <  \left(T_c\right)_q= \frac{h^2}{\left(2\pi\right)^{2/3} 2mk}
\left(\frac{N}{2.315 V}
\right)^{2/3}  \left( 1 + (q-1) 0.886 \right),
\eeq
where $\left(T_c\right)_q$ is the $q$-dependent characteristic
temperature of the Bose-Einstein condensation. It is easily seen that this
result shows that the critical temperature decreases when $q<1$, which
is consistent with the previous result of Curilef \cite{curilef}.
Note that the standard
$\left(T_c\right)_1$ case can easily be obtained for $q=1$ value in the
above expression.

%%%%%%%%%%%%%%%%%%%%%%%%%%%%%%%%%

Any accurate simultaneous determination of $N$, $V$ and $T$
can yield, in principle, a bound upon $q$. However in practice, this
could
be well below any practical possibility, due to the smallness of the
correction term.

\section{Conclusions}

Our main results can be summarized as follows. We have been able to solve,
using the generalized quantal distribution functions in the factorization
approach, the values for the average number of particles and energy
in the case of system of non-interacting particles, either fermions or bosons.
These results are expected to be useful in any analysis of statistical
phenomena
within the context of non-extensive scenarios.  We
could explicitly see that all the terms coming from the non-extensive part of
the integrals go to zero when the temperature goes to zero, something
that could
be expected due to the form of $n_q$.
Non-extensivity
can not play a role for very low temperatures, at least up to order $(1-q)$.
As a consequence, for instance,
the well known
Chandrasekhar limit for white dwarfs stars is not affected by a change of
the statistical framework. However, it is to be explicitly stated that any
other model of star, that happens with $T \neq 0$, will be affected by such
change. This is why, for example, this statistical scenario could be useful
to tackle the solar neutrino problem \cite{solar}.
In the boson case, we have seen that a small correction appears to the
Bose-Einstein condensation, this happening not exactly at $T=0$.
If this correction is enough to work out a possible bound upon
$q$ in this system remains to be studied. In the paper by Curilef it was argued
that due to a possible fractility of the universe, distribution functions could be
modified in the sense described here, and that this could be useful to study the
behavior of diffuse gas clouds. This could be an example of where the $q \neq 1$
statistics may arise, although due to the form in which equilibrium distributions
of boson stars arise it is unlikely that this could be applied for those cases.

%%%%%%%%%%%%%%%%%%%%%%%%%%%%%%%%%%%%%%%%%%%%%%
\subsection*{Acknowledgments}
During the course of this research, D.F.T. was a Chevening Scholar and
has been supported by the British Council (UK) and CONICET (ARGENTINA).
He would like to thank R. Borzi for valuable advice.
U.T. is a TUBITAK M\"{u}nir Birsel Foundation Fellow and
acknowledges partial support from Ege University Research Fund
under the Project Number 97 FEN 025.
We thank an anonymous referee for kindly suggesting
the alternative method to solve the integrals commented
above, among other useful insights.
%%%%%%%%%%%%%%%%%%%%%%%%%%%%%%%%%%%%%%%%%%%%%%%

%\newpage

\newpage

%\vspace{2cm}

{\bf Figure Captions}

\vspace{1.5cm}

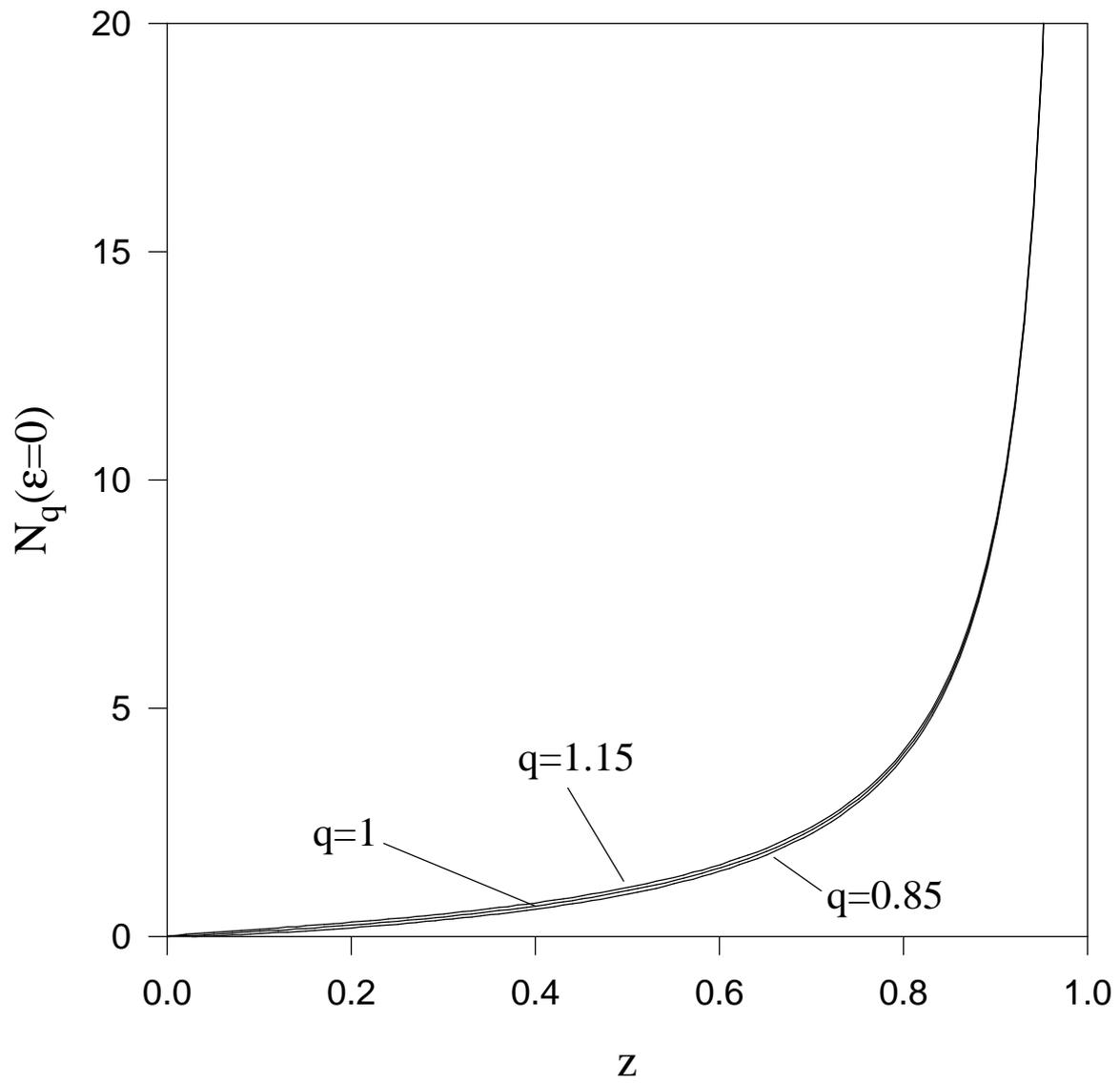
Figure 1 : The plot of $N_q(\epsilon=0)$ as a function of $z$ for various $q$
values.

\vspace{1cm}

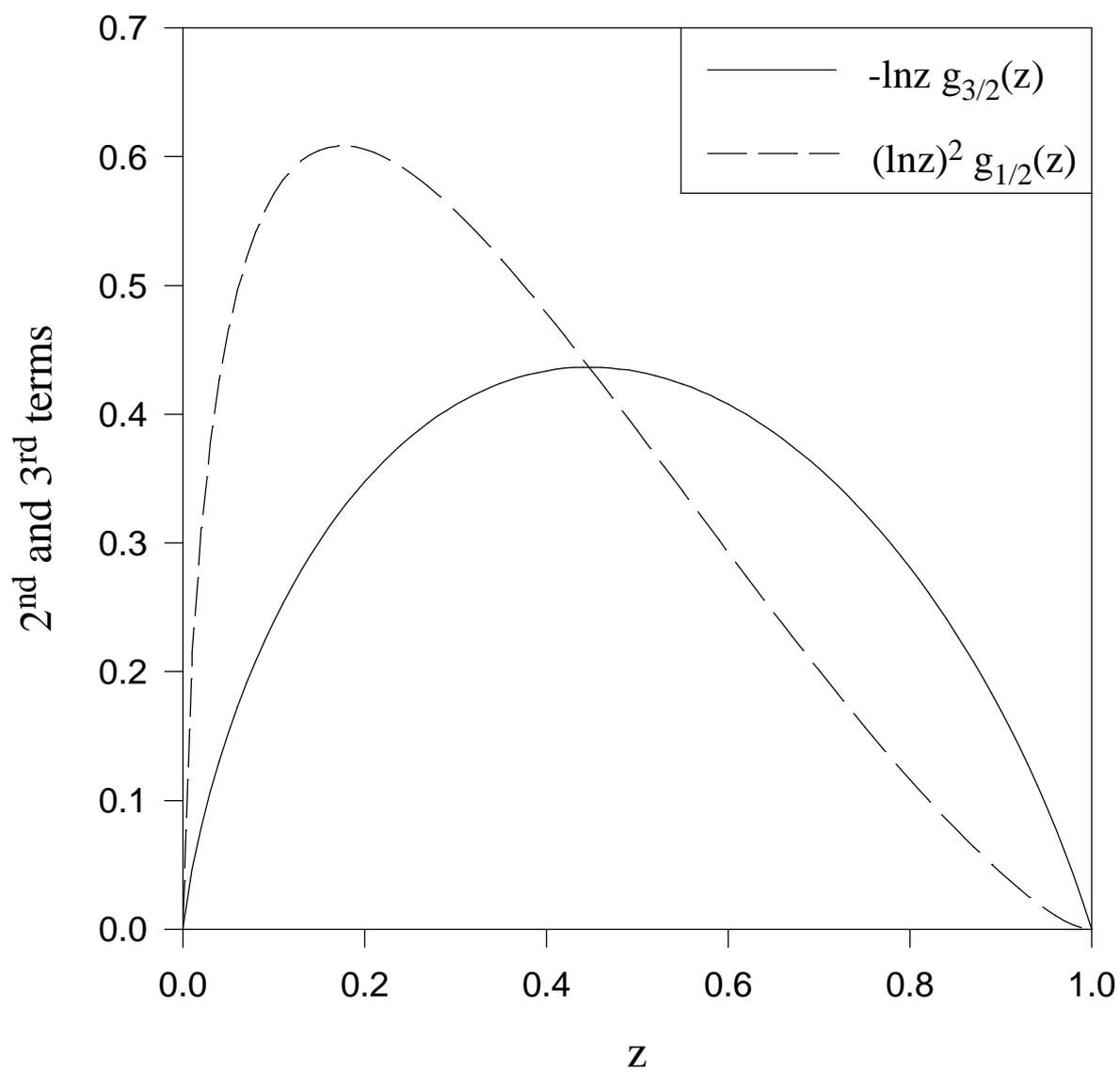
Figure 2 : The plot of the second and third terms of the $q$-dependent part
of eq.(43) as a function of $z$.

\end{document}